\documentclass{article}
\usepackage{sao2}

\usepackage{graphicx}
\usepackage{amssymb}
\usepackage{url}

\issue{2006, 59, 5-17}
\title{Collimated high-energy photons and other possible
observational effects of the photon angular and spectral
distribution in gamma-ray burst sources
}
\author{V.V. Sokolov\inst{a}\thanks{Email: sokolov@sao.ru (VVS)}
          \and V.G. Kurt\inst{b}
                \and G.S. Bisnovatyi-Kogan\inst{c}
                      \and Yu.N. Gnedin\inst{d}
                            \and Yu.V. Baryshev\inst{e}
                              \and T. A. Fatkhullin\inst{a}
                                    \and V.S. Lebedev\inst{a}
                                }
                                \institute{
                                \saoname
\and
Astro Space Center of the Russian Academy of Sciences, Moscow 117810, Russia
\and
Space Research Institute, Moscow 117810, Russia
\and
Central Astronomical Observatory at Pulkovo, Saint-Petersburg 196140, Russia
\and
Institute of Astronomy, St-Petersburg State University, Staryi Peterhoff 198504,
Russia
}

\def\g{$\gamma$}
\begin{document}
\date{October 25, 2004}{January 10, 2005}
\maketitle
\begin{abstract}
Typical observational gamma-ray burst (GRB) spectra are discussed
and, in this connection, what is the origin of the compactness problem
and how it was solved at first.
If the threshold for $e^{-}e^{+}$ pair production depends
on an angle between photon momenta, then another solution 
of the compactness problem is possible.
We discuss a possibility of the $\gamma$-rays collimation
and the dependence of photon beaming on photon energies.
The list of basic assumptions of the scenario describing the GRB
source with energy $< 10^{49}$\,ergs is adduced:
the matter is about an alternative to the ultrarelativistic fireball
if {\it all} long-duration GRBs are related or physically connected
with {\it normal/unpeculiar} core-collapse supernovae (SNe). 
Namely, we consider the questions about
radiation pressure and how the jet arises on account of
even small asymmetry of the radiation field in a compact GRB source.
The possibility of a new approach to explanation of the GRB phenomenon is shown.
Possible mechanisms of their generation in regions of size $< 10^8$cm are
discussed (a compact model of GRBs).
Observational consequences of the compact GRB energy release are considered.

\end{abstract}

\section{Introduction}

There are direct and indirect observational arguments
in favor of physical connection between massive
or core-collapse supernovae (SNe) and long duration
gamma-ray bursts (GRBs), and the list of publications
on the topic is ever-increasing. At first this connection
was justified by the fact that all GRB host galaxies
turned out to be star-forming or star-bursting
galaxies with high massive star-forming rates (e.g.,
Djorgovski et al. 2001; Frail et al. 2002; Sokolov
et al. 2001). There are more and more
occurences of SN signs in the GRB afterglow light
curves and spectra for GRB 970228 (Galama et
al. 1999), GRB 970508 (Sokolov et al. 1998), GRB
980326 (Bloom et al. 1999), GRB 990712 (Bjornsson
et al. 2001), GRB 991208 (Castro-Tirado et
al. 2001), GRB 000911 (Lazzati et al. 2001), GRB
011121 (Bloom et al. 2002), GRB 021211 (Della Valle
et al. 2003), GRB 030329 (Hjorth et al. 2003; Stanek
et al. 2003), GRB 031203 (Thompsen et al. 2004).
  A comprehensive analysis
  of SN light in GRB afterglows
  has been recently done
  by Zeh, Klose, and Hartmann (2004),
  and see references therein.
If there were more cases of clear and indisputable
spectral and photometric signs of association between
normal core-collapse SNe (Ib/c type and others)
and GRBs, it would be a direct proof of the connection
between GRBs and massive stars.
The increasing statistics of the GRB-SN associations
can impose direct and strong observational constraints
on GRB beaming and, hence, we could have an {\it observational}
estimation of a real total energy reservoir of GRB sources.

The purpose of this paper is to describe the basic assumptions 
of a scenario of a GRB source with energy $< 10^{49}$\,ergs:
the matter is about an alternative to the ultrarelativistic fireball
if {\it all} long-duration GRBs are 
physically connected
with {\it normal/unpeculiar} core-collapse supernovae (SNe).
In Section 2 it should be discussed the typical GRB spectra,
which is the origin of the compactness problem, and how it was solved
at early stages of GRB studies. 
Sect. 3 concerns with another attempt of  solving the compactness problem,
namely, the dependence of the threshold for $e^{-}e^{+}$ pair production 
on the angle between photon momenta,
a soft collimation and the dependence of this collimation on 
GRB photon energy.
In Sect. 4 we discuss and justify observably 
a rather strong collimation of GRB radiation
(for a small number of hard photons in GRB spectra)
reaching near-earth detectors. 
In Sect. 5 we consider the radiation
pressure and how a jet or bullet arises if the reason of
the relativistic jet is a powerful light pressure
of the collimated/non-isotropic prompt radiation of the GRB source.
In section 6 we are examining in outline
some  possible compact mechanisms of the GRB phenomenon
in a compact GRB model.
Sect. 7 (Concluding remarks) concerns observational consequences of
the ``compact" GRB energy release, and
what new we will see in the sky if the compact GRB source model
is true indeed: particulary,
we would like to adduce in outline some re-analysis 
of the observational results of soft X-ray flashes (XRFs),
GRBs with strong X-ray excess in spectra (X-Ray-Rich GRB = XRR GRB), 
normal or classical GRBs, obtained at first with BeppoSAX
and then with HETE-2
(Amati et al. 2002; Lamb et al. 2003a).

So, we will try to understand the soft (in the meaning of photon energies)
observational GRB spectrum without involving huge kinematical motions 
of the radiating plasma {\it a priori}, or without an enormous 
Lorentz factor, $\Gamma \gg 1$.

\section{On the typical GRB spectra and typical photon energies}

The rapid temporal variability, $\delta T \sim 10$ msec,
observed in GRBs implies {\it compact} sources with a size smaller
than $c \delta T \approx 3000$ km. But here a problem immediately
arises for distant GRB sources (e.g. Carrigan and Katz 1992): too
large energy ($>10^{51}$ ergs) is already released in only soft
$\gamma$-rays ($<511$ keV and up to 1 MeV) in such a  small volume
for the sources at cosmological distances ($>1$ Gpc).
For a photon number density 
$n_{\gamma} \sim (10^{51}$ergs$/(m_e c^2))/(c \delta T)^3
\sim 10^{57}/(3000 km)^3 \sim 10^{32}$cm$^{-3}$ two
$\gamma$-ray photons with a {\it sum energy} larger than $2 m_e c^2$
could interact with each other and produce electron positron
pairs. The optical depth for pair creation is given approximately
by $\tau_{e^+e^-}\sim  n_{\gamma} r_{e}^2 (c \delta T) \sim
10^{16}$, where $r_e$ is the classical electron radius
$e^2/(m_ec^2)$
(the cross-section for pair production is $\sim r_{e}^2$ or
 $\sim 10^{-25}$cm$^2$
at these semirelativistic energies). It is the essence of 
a so-called ``compactness problem":
the optical depth of the relatively low energy photons
($\sim 511$\,keV)
would be so large that these photons could not be observed.

1. Usually in this definition a role of
the {\it high-energy photons} is emphasized: 
the $\gamma$-ray photons with energies larger than $2
m_e c^2$ (and $\gg 1$\,MeV) could interact with lower energy photons
and produce electron-positron pairs (Piran 1996; 1999; 1999a; 2004).
The average optical depth for this process is $\tau_{e^+e^-}\sim
10^{15} (E/10^{51}$ergs)$(\delta T/10$ msec$)^{2}$ for a typical
total GRB energy release $E \sim 10^{51}$ ergs in a small volume
(e.g. Piran 1999). The ``heavy"/hard (or high-energy) photons
are present in
observational GRB spectra as high energy tails which contain a
significant amount of energy. So, according to Piran, the
compactness problem arises because the observed spectrum contains a
fraction of the high energy $\gamma$-ray photons.
In other words, since observations are consistent with 
{\it a possibility} that all GRBs have the high energy tails,
it must be the first and basic observational justification of the problem
(see e.g. Lithwick and Sari 2001): the optical depth of the
high-energy photons ($\gg 1$\,MeV) would be so large that these photons
could not be observed. 
However, here we must make right away several
specifying remarks on the typical spectra and typical photon
energies in GRBs, including  allowance for the well known results
of recent observational spectra.

Yes, such high-energy photons did were observed in some cases,
but far from being so always. Furthermore,  photons with energies
 $>100$ MeV were strongly delayed after the main GRB burst.
For example, the 20 GeV photon observed with BATSE/EGRET delayed as much as
1.5 hours relative to the GRB itself. It is obvious that in this case
a physical mechanism was quite different from one creating typical
prompt GRB spectra.
(The GRB spectra are described in a review by Fishman and Meegan (1995),
see also the catalogue of the spectra by Preece et al. (2000).)
Typical observational GRB spectra turned out to be very diverse,
but yet these are mainly soft (but not hard) gamma-ray quanta.
It has been known  since the moment of GRBs discovery,
when their spectra were presented in energy units:
e.g., see a review by Mazets and Golenetsky (1987).
At present many authors point to the same again
(Lamb et al. 2003; Baring \& Braby 2004;
Liang et al. 2004; Atkins et al. 2003; Gialis \& Pelletier 2004).
Almost all GRBs have been detected in the energy range
between 20 keV and 1 MeV. {\it A few} have been observed above 100 MeV.
In a recent review Piran (2004) also has paid attention
to a puzzle of the origin of narrow
distribution for the typical energy
of the observed GRB radiation ($E_p < 511$\,keV, Preece 2000).
Besides, by 2000 it was clear that there were other two GRB classes:
X-Ray Flashes (XRF) and X-Ray Rich Gamma Ray Bursts (XRR GRB)
(Heise et al. 2001; Amati et al. 2002).
These are GRBs either {\it without} (XRFs) or almost without (XRR GRB)
gamma-ray quanta.
It is also discussed in detail and there are excellent illustrations
in recent papers
by Lamb et al. (2003a, 2003b, 2003c) and in other papers by this group.

     Thus, 
despite the importance of
the problem of the high energy ($\gg 1$\,MeV) photons release,
still there are too many lower energy $\gamma$-ray
photons in a small volume with $R\sim 3000$\,km.
The observed fluxes give an estimate
of a total GRB energy release to be of $\sim 10^{51}$ ergs
in the form of
just these {\it low energy} photons,
or this ``standard" estimation ($\sim 10^{51}$ ergs)
was obtained from
typical observational GRB spectra of just these,
most frequently observed low-energy (``target") photons with
the {\it semirelativistic} energies, up to 1 MeV, basically.
(It is natural that the photon
density was estimated using the simple assumption of spherical symmetry.
See below the comments to the paper by Carrigan and Katz 1992.)

2. Further it was firmly declared, and other authors have repeated
many times that GRB source must be optically thin
and the observed {\it spectrum is non-thermal} with certainty
(Piran 1996; 1999; 1999a; 2004).
Now the optically thin source
with the non-thermal spectrum
is presented as a standard
common opinion about all GRBs (Postnov 1999),
though unlike time-averaged GRB spectra, time-resolved
instantaneous  GRB spectra are thermal
({\it black-body radiation} with temperature $kT\sim 100$\,keV) rather
than power-law ones
(Crider et al. 1997; Preece et al. 1998, 2002; Ghisellini 2003;
Ghirlanda et al. 2003; Ryde 2004).
So far different authors have been
pointing to these inconsistencies between the standard optically thin
synchrotron model and observations (e.g. Preece et al. 2002),
and suggesting different alternative scenarios of the
solution of this problem (Blinnikov, Kozyreva and Panchenko 1999; Medvedev
2000; Baring \& Braby 2004).
It turned out that the black-body radiation with $kT\sim 100$\,keV
is a physical model while the time-averaged
non-thermal GRB spectrum
is merely an empirical model
(Ryde 2004, and other references therein).

Nevertheless, if these
 theoretical rather than observational statements (Piran 1996)
 on the possibility that {\it all} GRB spectra have
 high energy tails (1) and
 the observed GRB spectra are non-thermal (2),
are true indeed,
the fireball theory (Piran 1999, 1999a) with huge Lorentz factors
is the only possible theoretical alternative.
It should be admitted though that
the standard optically thin synchrotron shock emission theory/model
explains everything, except the observational spectra of GRBs themselves
(Preece et al. 2002).
But for all that, it was left out of account that these ``target-photons"
($E_p < 511$\,keV) are just the observed typical GRBs.
So, it turns out that the main task, according to the standard model,
is not the
explanation of this observed soft GRB spectrum in terms of photons'
energy/frequency, but the investigation of rare cases of  release
of hard quanta with energy of more than or $\sim 1$\,GeV.
In this connection see the paper by Lithwick and Sari (2001) in
which, as an alternative to the observed GRB spectrum,
 an ``intrinsic spectrum" that has no cutoff at very high
energies is suggested to be explained.

As a result, the origin of the {\it observed} and substantially soft GRB
spectra with a big number of photons up to $\sim 1$\,MeV remains
not properly understood. It is especially incomprehensible against
the background of conjurations about the huge gamma factor that is supposed
to solve the compactness problem. But the question remains: why are
mainly soft GRB spectra  observed at ultrarelativistic motions
of radiating plasma supposed in the fireball model? And what is
more, sometimes the GRB spectra do not contain \g-ray quanta at
all, as, for example, XRFs known already before 2000 (Heise et
al. 2001). Thus, when solving the compactness problem, we somehow
imperceptibly incurred another problem of strong contradiction
between the ultrarelativistic Loretz factor $\Gamma \sim$ 100-1000
(with 100 MeV and 10 GeV photons) and observed soft ($\sim$ or $<1$
MeV) \g-ray (GRB, XRR GRB) and X-ray (XRF) radiation of the most
classical GRBs. Moreover, it is also important to point out here
that the observed {\it black-body} prompt GRB radiation with a 
temperature $kT\sim 100$\,keV
(Ghirlanda et al. 2003; Ryde 2004)
is inconsistent with the Loretz factor $\approx 10^2 - 10^4$ for the
reason that the mean observed temperature can easily exceed MeV in
cosmological fireballs (Piran and Shemi 1993).


\section{The threshold for $e^{-}e^{+}$ pair production 
         depends on the angle between photon momenta}

So, is there the compactness problem or not?
If yes, then how is it solved? Are there any alternatives of its solutions
besides the fireball with its huge Lorentz factor?
Particularly, can we do with semi- (not ultra-) relativistic approximation
when explaining observed GRB, XRR GRB and XRF spectra?
Is the strong gamma-radiation beaming necessary and to what extent
can the radiation in GRB, XRR GRB and XRF spectra be collimated?
This section concerns another attempt to solve the 
old compactness problem.

Certainly, in 1998 there already were some discussions of radiation
collimation, but mainly in terms of the same standard fireball theory.
And it should be kept in mind
that in this theory the term {\it collimation} refers to jets consisting
of plasma, while the term {\it beaming} refers to radiation of the same
optically thin plasma (Sari 2000).
Of course, we could waste not much time for the discussion of Piran's
approach in the previous section,
if it were not a circumstance that even before 1992
(i.e. before the BATSE/EGRET mission) the compactness problem was mentioned
in connection with the famous burst of 1979 March 5 in the
Large Magellanic Cloud. Already then a possibility of collimated
$\gamma$-ray radiation in explanation of observed soft spectra
was not excluded because
the cross-section of electron-positron pair production
$\sigma_{e^-e^+}$ (and annihilation also) depends not only on energy,
but on the angle between momenta of colliding particles.

Then there is a comment on
the paper by Carrigan and Katz (1992) which has not been so often cited.
In fact as early as at the beginning
of the 1990th a lot of interesting was said in connection with collimation
of $\gamma$-rays leaving the source with high photon density in it.
It seems that just the collimation solves the problem (see below).
The paper by Carrigan and Katz (1992)
tells about modeling the observed GRB spectra allowing for the
electron-positron pair production effects.
These effects could produce effective collimation of the flux
because of kinematics of the two-photon pair production:
the opacity ($\tau_{e^-e^+}$) is also a sensitive
function of the {\it angular} and {\it spectral} distribution of the
radiation field in the GRB source.

Because of the importance of the photon
angular and spectral distribution to the opacity,
below an analysis of formula (1) for
the threshold of the pair production
processes from the paper by Carrigan and Katz (1992) is given.
The argument proceeds as follows:
{\it two} photons with energies $E_1$ and $E_2$,
which are above the threshold
energy ($E_1+E_2 > 2\cdot E_{th}$) for electron-positron pair production


\begin{equation}
E_1 \cdot E_2 \geq 2(m_e c^2)^2 / (1-cos\theta_{12})
\end{equation}
may produce a pair, where
$2(m_e c^2)^2 = 2(511 keV)^2$,
$\theta_{12}$ is the angle between the directions
of the two $\gamma$-rays,
and $E_{th}= \sqrt {E_1 E_2}$.
The cross section for pair production reaches
the maximum at a finite center-of-momentum photon energy:
e.g. $E_1 + E_2 > 2\cdot E_{th} = 2\cdot$511~keV for $\theta_{12} = 180\degr$,
or $E_1 + E_2 > 2\cdot E_{th}\approx 2\cdot$700~keV
for $\theta_{12}\approx 90\degr$),
or $E_1 + E_2 > 2\cdot E_{th}$ tending to infinity ($\gg$1~MeV)
for $\theta_{12}\approx 0\degr$.
  If the source photon spectrum is not sharply peaked,
the relatively high-energy photons ($E > E_{th}$) will,
therefore, form pairs predominantly with relatively
low-energy photons ($E < E_{th}$).
It means that the observed/released GRB spectra will be soft, since the
high-energy photons will be held by the threshold of pair production.
Thus, because any {\it reasonable} source spectrum will contain much more
low- or  moderate-energy photons ($\lesssim 511$\,keV)
than high-energy photons,
the emergent spectrum will differ most markedly from the source spectrum
at high photon energies ($E\gtrsim 1$ MeV)
at which it (the emergent spectrum) will be heavily depleted.
In other words, the observed (emergent) spectrum becomes softer.
Then, the $e^{-}e^{+}$ pairs eventually
annihilate to produce two (infrequently 3) photons,
but usually not one high- and one low-energy photon.


The result is that high-energy photons are preferentially removed from the
observed spectrum. The observation of a measurable amount of flux with
$E > E_{th}= \sqrt {E_1 E_2}$
is not expected unless the optical depth $\tau_{e^-e^+}$
to pair production is equal to 1 or less,
because the threshold for electron-positron pair production (1)
is also a sensitive function of the angular distribution of the
radiation field (in the very source).


Thus, the observation of a considerable number of quanta with $E > 1$ MeV
due to the filter effect (1) is not expected, if only the optical depth for
the $e^{-}e^{+}$ pair production is not proved $\lesssim 1$ indeed.
As is seen from the paper by Carrigan and Katz (1992), in 1992 it was generally
accepted that typical energies of most photons in {\it observed} GRB spectra
are still rather small. Further in the peper, Carrigan and Katz adduce
the estimates of distances to burst sources of such photons
with the {\it semirelativistic} energies.
The matter is that the problem of a compact source
(in relation to the 1979 March 5 event in LMC)
and a surprisingly big distance arises indeed.
But not because of 
a problem with the release of
``heavy" (100 MeV, 1 GeV, or more)
ultrarelativistic photons which interfere with ``light" ($\lesssim 1$ MeV)
target-photons observed in the GRB spectra.
The powerful 1979 March 5 event in LMC was observed without any
super heavy photons in its spectrum.
To make sure of it
one should just look at the spectra of this
burst published by Mazets and Golenetskii in their review (1987).

To explain why the effect of the photon ``$e^{-}e^{+}$ confinement"
  does not function in
this GRB source (1979 March 5 event in LMC), Carrigan and Katz discuss
different possibilities. In particular, they immediately point
out to the angle dependence (1) of the threshold
of the $e^{-}e^{+}$ production.
A possible ``loophole" exists if the source produces a {\it strongly
collimated} beam of photons.
(Thus, the question is about an asymmetry of the radiation field
in the source.)
In this case, even high-energy photons are below the 
threshold for the pair production if $\theta_{12}$ is small enough.
The presence of such a ``window" in the opacity for {\it collimated} photons
suggests that in a region opaque to pair production
much of the radiation may emerge through this window,
in analogy to the great contribution of windows in the material
opacity to radiation flow in the usual (Rosseland mean) approximation.

The use of the words ``strongly collimated" in this (``old") paper could be
somewhat confusing. What means {\it strongly} indeed?
At that time there were no observations of GRB spectra in the region
of high energy $E$.
Heavier photons with $E \sim 10$ MeV (beyond the peak of $\sim 1$ MeV)
have been reliably observed only with EGRET/BATSE.
In particular, from formula (1) for such photons an estimation of the
collimation angle can be obtained (without any ``target-photons"):
$1 - cos\theta_{12} = 0.522245 MeV^2/(10MeV \cdot 10MeV) \approx 0.005.$
It corresponds to $\theta_{12} $ less than 6 degree only.
It means that
the quanta with energy $\sim 10$MeV leaving the source within
a cone of $\sim 6^o$ opening angle
do not give rise to pairs, and all {\it softer} radiation
can be uncollimated at all. So the collision of 10 MeV quanta
with quanta of lower energy occurs at angles greater than
($0.522245 MeV^2/(10MeV \cdot 100KeV) \approx 0.5$) $60^o$, and softer
quanta leaving the source within the cone of such opening angle do not prevent
neither heavy nor (especially) light quanta to go freely to infinity.

Thus, formula (1) demands more or less strong collimation only for {\it
a small part} of the heaviest quanta radiated by the source. If one looks at
energetic spectra of typical GRBs (the same reference to Mazets and
Golenetskii 1987) presented in the old way of
$F(cm^{-2} s^{-1} KeV^{-1})$~vs.~$E(KeV)$ --- {\it the number} of photons
per a time unit in an energy range unit per an area unit versus the photons
energy, --- then everything becomes clear. Only a small part or a small
{\it amount} of quanta/photons observed beyond a threshold of
$\approx 700$KeV can be collimated,
but within a cone of $< 90^o$ opening angle :).

At present, 6 degrees for 10 MeV quanta would not be considered as a strongly
collimated beam. Now such opening angles (of jets)
are considered to be quite suitable in the ``standard"  or the most
popular theory of fireballs.
If one proceeds
right away from an idea that it is necessary to release
quanta with the energy up to 10 MeV, then we
would obtain at once
a version of a collimated theory with the $\Gamma$ of $\sim 10$.
But such a way in the standard fireball theory is a dead end also.
The allowing for an initial collimation of GRB radiation
can drastically change
this model (see below) for the collimation arising
{\it directly in the source}
but not because of a huge $\Gamma$ of $\sim 1000$
what would be needed to solve the compactness problem.

One way or another, the light flux is to lead to corresponding effects of
radiation pressure upon the matter surrounding the source. 
And if in addition the radiation is collimated, then 
the arising of jets (at so enormous light flux)
becomes an inevitable consequence of even a small
{\it asymmetry} of the radiation field in the source.
But the question is if

\section{Is the jet a GRB source or not?}

Indeed, perhaps one should take into account right away this angular
dependence of the threshold of the
pair $e^{-}e^{+}$ production (1) before
the ultra relativistic limit, allowing for a possibility of a preferential
(most probably by a magnetic field) direction in the burst source on the
surface of a compact object -- the GRB source. Does a preferential direction
in the source sound wrong? But one way or another, in the model of fireball
with jets the radiating plasma is to be accelerated up to enormous
velocities. What is the mechanism? In the fireball theory this question is
not solved yet, and the origin of GRB spectra also remains incomprehensible.
In the end, does the jet radiate itself and is it the GRB source? That is
the question. Can we do without the radiating and accelerated
(nobody knows by what) jet up to a huge value of the Lorentz factor,
by supposing that the source of GRB radiation is {\it already} collimated
by the burst source itself?
At least, the rather strong collimation of GRB \g-rays,
reaching near-earth detectors, can be observably justified.
The GRBs could be the beginning of the explosions
of {\it usual} massive or core-collapse SNe
 (Sokolov 2001a, 2001b).

All results of photometrical and spectral observations of host galaxies
confirm the relation between GRB and evolution of a massive star, i.e.,
the close connection between GRB and 
relativistic collapse with SN explosion in the end of the star evolution.
(Here it is already possible to adduce a lot of references:
Djorgovski et al. 2001; Frail et al. 2002; Sokolov et al. 2001; etc.)
The main conclusion resulting from the investigation of these galaxies is
that the GRB host galaxies do not differ in anything from other galaxies
with close value of redshifts $z$: 
neither in colors, nor in spectra, star-forming rates, 
luminosities, and surface brightness. 
{\bf It means that these are the galaxies (``ordinary" for their redshifts)
constituting the base of all deep surveys.}

In point of fact, this is the main result of
{\it optical identification} of GRBs with (ordinary) objects of already known
nature: GRBs are identified with galaxies up to $\approx 26$ st. magn.
With allowing for the results of direct optical identifications
this makes it possible to estimate directly from observations
an average yearly rate of GRB events in every such galaxy
by accounts of these galaxies for the number of galaxies brighter than
26th st.
magn. It turns out to be equal to
       $ N_{GRB} \sim 10^{-8} yr^{-1} galaxy^{-1}$.
(But most probably this is only an upper estimate, see in Sokolov 2001b).
Allowing for the yearly rate of (massive) SN explosions
       $ N_{SN} \sim 10^{-3} - 10^{-2} yr^{-1} galaxy^{-1}$,
the ratio of the number of GRBs, related with the collapse of massive stars
(core-collapse SNe), to the number of such SNe is close to
$ N_{GRB}/N_{SN} \sim 10^{-5} - 10^{-6}$.
Most likely, this is also only the upper estimate for Ib/c type SNe
(Sokolov 2001a). Porciani \& Madau (2001) obtained an analogous estimate:
$ (1 - 2)\cdot 10^{-6}$ for II type SNe.

Here we proceed from the simplest assumption, which has been confirmed from 1998
by increasing number of observational facts, that {\it all} long-duration
GRBs are related to explosions of massive SNe.
Then the ratio $ N_{GRB}/N_{SN}$
should be interpreted as a very strict ``$\gamma$-ray beaming" of quanta
{\it reaching an observer}, when gamma-ray radiation (a part of it) of the
GRB source propagates to very long distances within a very small solid angle

\begin{equation}
     \Omega_{beam} = N_{GRB}/N_{SN} \sim (10^{-5} - 10^{-6})\cdot 4\pi.
\end{equation}

Another possible interpretation of the small value of
$ N_{GRB}/N_{SN}$ --- a relation to a rare class of some
peculiar SNe --- seems to be less possible (or hardly probable),
since then GRBs would be related only to the
$10^{-5}-10^{-6}$th part of all observed SNe in distant galaxies
(up to 28th mag).
These are already not simple peculiar SNe, with which the Paczy\'nski's
hypernova is sometimes identified (Paczy\'nski 1999; Fields et al. 2002).
Peculiar supernovae/``hypernovae", such as 1997ef, 1998bw, 2002ap, turn out
to be too numerous
(Richardson et al. 2002; Podsiadlowski et al. 2004)

Now there are already other papers 
(Lamb et al.
2003a, 2003b, 2003c),
pointing out to a possibility of collimated radiation 
from the GRB source (2).
And the more numerous are GRB/SN coincidences of type of
GRB 030329/SN 2003dh or GRB/``red shoulder" in light curves, the more
confident will be the idea that GRB radiation is collimated, but not
related to a special class of SNe/``hypernovae".
Many consider this term ("hypernovae") poorly defined and no longer use it.
The more so, that explosion geometry features 
 (SN explosion can be axially symmetrical)
make the attempts to select a class of ``hypernovae" more complex
(Willingale et al. 2004, see the end of their text).
Today there are more facts for the collimation (2), 
and we think that soon it will be accepted not only
by Lamb et al.

Let us suppose that only the most collimated part of gamma radiation get to an
observer, say, along a rotation axis of the collapsing core of a star with
magnetic field.
And if GRBs are so highly collimated,
radiating only into a small fraction of the sky,
then the energy of each event  $E_{beam}$
must be much reduced, by several orders of magnitude
in comparison at least with a (so called) 
``isotropic equivalent" $E_{iso}$, of a total GRB energy release
  ($E_{iso} \sim 10^{51} - 10^{52} ergs$ and up to $10^{53} ergs$):
\begin{equation}
     E_{beam} = E_{iso} \Omega_{beam}/4\pi \sim 10^{45} - 10^{47} ergs .
\end{equation}

If it is just this case which is realized, and if the energy of \g-rays
propagating in the form of a narrow beam reaching an observer on Earth
is only a part of the total radiated energy of the GRB source
(from $\sim 10^{47} ergs$  to $\sim 10^{49} ergs$), then the other part
of its energy can be radiated in {\it isotropic} (or almost isotropic) way
indeed.
But at the spherical luminosity corresponding to a total GRB energy of,
e.g., $\sim 10^{45} - 10^{47} ergs$, no BATSE gamma-ray monitor detector,
even the most sensitive one, would detect flux, corresponding to so low
luminosity for objects at cosmological distances of $z \gtrsim 1$, and if the
observer is outside the cone of the collimated component of radiation (2).
I.e. (3) can be close to the lower estimate of the total radiated energy
of GRB sources, corresponding to the flux measured within the solid angle
(2), in which the most collimated component of the source radiation is
propagating. (We always suppose that {\it all} long-duration GRBs
are related to SNe.)

So, in terms of observational results known today, there is a possibility at
least to considerably reduce at once the total (bolometric) energy of GRB
explosions even in the model with radiating plasma, i.e. with ultra
relativistic jets in the standard fireball model.
(Though we are sure that it is not plasma that
 radiates the GRB, and the jet is a consequence, but not
the cause of GRBs. More will be said below.)
Then it is possible to estimate the Lorentz factor $\Gamma$
in the same standard theory
with the radiating jet (the formula and relation with 
the $\Gamma$ are taken from the paper by Piran (1999).
Even in the theory with the jet radiating GRB,
but at the total energy of
$10^{45}, 10^{47}, 10^{48}, 10^{49}$ ergs, the Lorentz factor 
$\Gamma$ turns out to be equal to: 18   ,  32  ,  42  ,  56  correspondingly.
But we do not think the authors of the standard solution of the compactness
problem will ever agree to that,
though here it is possible to speculate using the closeness of this estimate
to what was mentioned above for the
angle of collimation and the factor $\Gamma$ of photons with energy of
$\sim 10$\,MeV (see the end of the previous section). 
Maybe, Lamb and his co-authors (2003a, 2003b, 2003c)
will do so, since they try to adjust the very small angle of the GRB collimation
with the standard theory of the radiating jet.
But in our opinion, there is only one
alternative for the approach (Lamb et al. 2003c):
the model (``A Unified Jet Model of X-Ray Flashes, X-Ray-Rich GRBs, and
GRBs") does explain observations, but with $\Gamma$ of $\sim$1 - 10.
Then both the opening of jet and the angle of the GRB collimation
are to be simply
equal to each other for sure, and the GRB source is to be located in
the very beginning, or in the ``point" where the jet and radiation arise
(see Fig. 5. b. in the paper by Lamb et al. 2003c). But this will be quite
a {\it non-standard} theory.

Apparently, this question -- what does radiate:
a ``point" or an extent jet? -- is crucial for any
GRB mechanism. If the GRB source radiation (mainly the hard component of
the GRB spectrum) is collimated indeed,
then we will have to return to the old idea: the radiation (GRB) arises 
{\it on a surface}
(centimeters, meters?) of a compact object.
(Perhaps, the radiation in an annihilation line will be also found again.)
Further we will try to do without an (a priori) assumption that it is only the
jet's ``end" which radiates.
The jet is rather a consequence, than a cause. It arises for sure, but
because of the strong pressure of the collimated radiation on the matter
surrounding a compact (down to $10^7$ cm and less) GRB source. Certainly,
this jet accelerated by photons up to relativistic velocities will radiate
also, but it would be already an afterglow, but not GRB itself.

\section{The radiation pressure and origin of the jet in the compact model of GRB}

If the scenario: {\it massive star} ---$>$ {\it WR star} ---$>$
{\it pre-SN = pre-GRB} ---$>$ {\it the collapse of a massive star core} 
with formation of a shell around WR is true, 
then it could be supposed that the reason for arising
of a relativistic jet is the powerful light pressure of the collimated or 
non-isotropic prompt radiation of the GRB source onto the matter 
of the WR star envelope located immediately around the source 
itself --- a collapsing core of this star.

We can digress for a while from the problem of the mechanism of arising of
the GRB source itself and not discuss a question of how these collimated
gamma-quanta arouse. For example, the 
radiation field arising around the source can be
non-isotropic --- axially symmetric due to magnetic field and effects of
angular dependence (1) of the threshold of the
$e^{-}e^{+}$ pair production.
After all, for a while it is sufficient for us that only a part ($\sim 10\%$ or
even
$1\%$) of the total GRB energy ($\sim 10^{47} - 10^{49} erg$)
may be the collimated radiation, which breaks through the dense envelope
surrounding the collapsing core of the WR star.
(Then the prompt radiation reaches the Earth and is detected as the GRB.)
The main thing now is {\it the collimated flux} of radiation from the source and
a possibility of existence of {\it dense} gas (windy) environment
pressed up by radiation from the GRB source embedded in it, and this
environment can be the most dense just near the source, if the density is
close to $n = A r^{-2}$ (the WR law for stellar wind).
Here the distance $r$ is measured from the WR star itself, and
$A\sim 10^{34}$ cm$^{-1}$ (Ramirez-Ruiz et al. 2001).

For the force of light pressure that can act on gas environment
(plasma) around the GRB source (the WR star) we have
$L_{GRB} \cdot (4\pi r^{2})^{-1} \cdot (\sigma_{T}/c)$, where $L_{GRB}$ is a
so called isotropic {\it luminosity} equivalent of the source
($\sim 10^{50-51}$ erg$\cdot$ s$^{-1}$ and more),
$r$ is a distance from the center (or from the source),
$\sigma_{T} = 0.66 \cdot 10^{-24}$cm$^2$ is the Thomson cross-section, $c$ is
the velocity of light.
It is clear even without detailed calculation that near the WR core 
($r\sim 10^9$cm) such a force can over and over exceed (by 12-13 orders!) 
the light pressure force corresponding to {\it the Eddington limit} 
of luminosity ($\sim 10^{38}$erg$\cdot$s$^{-1}$ for 1 $M_\odot$).

In principle, the isotropic radiation with so huge luminosity
$L_{GRB}\sim 10^{50-51}$erg$\cdot$s$^{-1}$ 
(or the light pressure)
can also lead to fast acceleration
(similar to an explosion) of environment adjacent to the source.
But if we assume that the radiation of the GRB source is non-isotropic and a
part of it is collimated or we have very strong beaming with the solid angle
   $\Omega_{beam} \sim (10^{-5} - 10^{-6})\cdot 4\pi$,
then the forming of directed motion of relativistic/ultra-relativistic
jets becomes inevitable, only because of so huge/enormous light pressure
affecting the {\it dense} gas environment in the immediate vicinity of the
source - collapsing stellar nucleus.
Naturally, the formation of jets depends also on degree of ionization,
density and temperature of a medium in the immediate vicinity 
of the GRB source ---
an asymmetric collapsing nucleus of a massive star 
(Gorbatsky 2004, private communication).
But
we can estimate the size of the region {\it within} which such a jet can be
accelerated by the radiation pressure up to relativistic velocities: \\
1.~~~If the photon flux producing the radiation
pressure accelerating the matter at a distance $r$
from the center (near the GRB site) is equal to
   $L_{GRB}\cdot (4\pi r^{2})^{-1}$, 
then in the immediate vicinity from the GRB source
(the collapsing nucleus of WR star)
such a flux can be enormous.
It is {\it inside} this region where the jet originates
and undergoes acceleration up to
ultra relativistic velocities. \\
2.~~To accelerate the matter up to velocity of at least $\sim 0.3c$,
at the {\it outer} boundary of this region the photon flux must be at least not
less than the Eddington flux
$L_{Edd}\cdot (4\pi R_{*}^{2})^{-1}$.
Here $L_{Edd}$ is the Eddington limit
$\sim 10^{38}$erg$\cdot$s$^{-1}$  for 1 $M_\odot$  and $R_{*}$
is the size of a compact object of
$\sim 10^{6}$ cm.
 (By definition: $L_{Edd}\cdot (4\pi R_{*}^{2})^{-1}$
is a flux {\it stopping } the accretion onto a compact source
--- the falling of matter on the source at a parabolic velocity.
For a neutron star it is equal to $\sim 0.3c$.)

From the condition that the photon flux
$L_{GRB}\cdot (4\pi r^{2})^{-1}$ at distance $r$ is equal to
$L_{Edd}\cdot (4\pi R_{*}^{2})^{-1}$
(or at least not less than this flux),
and taking into account that the luminosity or rather its {\it isotropic
equivalent} of the GRB radiation is
$L_{GRB}\sim 10^{50-51}$erg$\cdot$s$^{-1}$,
it is possible to obtain an estimate of the size of
$\sim 10^{12}$ cm $\approx 14R_\odot$.
At least, at this outer boundary the light pressure is still able to
accelerate the initially stable matter up to sub-light velocities $\sim
0.3c$. 
And {\it deeper}, at less distances than $\sim 10^{12}$ cm from the
source, say, at $r\sim 10^{9}$ cm (somewhere {\it inside} the region of the
size less than the characteristic size of collapsing core of the massive
star) the light accelerates the matter up to ultra relativistic velocities
with the Lorentz factor of  $\sim 10$ at 
$L_{GRB}\sim 10^{50}$erg$\cdot$s$^{-1}$.
It can occur in a rather small volume of the typical size of
  $\lesssim R_\odot$,
which, in particular, agrees with observations
of the variable absorption feature
observed simultaneously with
GRB 990705 in its BeppoSAX/WFC spectrum (Amati et al. 2000).
Thus, inside the region of a size of less (in any case) than $10-15 R_\odot$,
a relativistic jet arises as a result of the strong light pressure onto the
ambient medium.

Certainly, the question about deceleration of such a jet in
circumstellar medium of the star progenitor should be considered
separately.
But perhaps the strong deceleration
due to interaction of the relativistic shock with ambient medium 
does not arise (as in the fireball model) even at very high 
densities of this matter around the WR star 
(up to $n\sim 10^{10}$ cm$^{-3}$ for $r\sim 15 R\odot$)
because the compact relativistic jet (or ``bullet") is decelerating but not
the shock, what is in the model by
Panaitescu (2001), Panaitescu \& Kumar (2001, 2002).
Here there is no such a wide ``bulldozer" --- a shock wave
raking up the matter 
and, correspondingly, 
there is no or almost no effective deceleration of ``the bullet"
as it moves towards less dense matter (with $n = A r^{-2}$) 
around the massive core of WR star.
That is why due to the small deceleration and small radiation losses 
(but with a large initial momentum),
this ``bullet" can move at a relativistic speed
{with the same Lorentz factor $\Gamma$ of $\sim 10$} all the time while 
the transient (or the GRB afterglow) is observed, 
i.e. over all its light curve with its peaks or breaks.
The shocks, which arise as the jet moves through, 
only heat this medium and then
are radiated in X ray, in optical, in radio where this medium is still dense
enough + non-uniformities in distribution of $n(r)$ at distances $ \sim
10^{15} - 10^{17}$ cm from the source (Sokolov 2001b).

Below is the list of basic assumptions of the scenario describing the GRB
source with energy of order $10^{47} - 10^{49}$\,ergs and non-empty space
near a massive star progenitor: \\
1.~~Around the WR star progenitor of GRB source,
from a distance of $\lesssim 10^{9}$\,cm 
(the typical size of a massive star core) up to $r\gtrsim 10^{15}$\,cm
(the distance where the interaction between WR wind and
ambient/circumstellar medium begins), 
there is a dense or windy medium --- an envelope 
resulting from the evolution of massive star. \\
2.~~The huge light pressure is the cause of the arising of the jet
in the region of $\sim 10^{9}$\,cm to $\sim 2\cdot 10^{11}$\,cm,
i.e. where the envelope density ($\sim 10^{15} - 10^{10}$\,cm$^{-3}$) 
is the highest, but the optical depth for Tompson scattering 
can be already less than 1 ($\tau \sim \sigma_{T} \cdot n \cdot r <1$). \\
3.~~The burst itself, probably an almost spherically symmetrical
``GRB-explosion", with a total energy up to $\sim10^{49}$\,ergs arises
somewhere
in the volume of size $\sim 3\cdot 10^{9}$\,cm,
or at even a smaller depth of $\sim 10^{8} - 10^{6}$\,cm, i.e. where the WR law
$n = A \cdot r^{-2}$ (for the stellar wind) ceases to be valid.
It is possible that
the explosion/burst occurs directly on the surface of a compact object,
resulting from massive star core collapse. \\
4)~ Only the most collimated radiation part of the GRB source
propagating within the solid angle of $\Omega_{beam} \sim (10^{-5} - 10^{-
6})\cdot 4\pi$\,sr goes to infinity, and for all that the total energy of the
source is either of the same order as $E_{beam} = E_{iso}
\Omega_{beam}/4\pi \sim 10^{45} - 10^{47}$\.ergs or about $10^{49}$\,ergs.

Now we can return again to the question about collimation of photons
arising on the surface of the compact object.
The pairs production will not prevent the photons with wave
vectors within a solid angle of the opening of $\theta_{12}$
for \mbox{$(1-cos \theta_{12}) < 2(m_e c^2)^2/ E_1\cdot E_2$} from free exiting to
infinity.
The threshold
\mbox{$E_{th}=\sqrt {E_1 E_2}$}
inside this solid angle at small
$\theta_{12}$ (for \mbox{$\sqrt {E_1 E_2 /2}\gg$ 511~keV)}
is very high, and all photons with energies
below $E_{th}$ will freely exit through this ``window" in
opacity for collimated photons.
An initial spectrum of the GRB source is almost what is observed,
still the filter
(1) affects the most violently the hard range of the observed GRB
spectrum --- it was mentioned above (Carrigan and Katz 1992).
And one should not try
``to invent" at once a special mechanism of a very sharp
collimation/``channeling" of all photons. It was also mentioned above.


Thus, it is undoubtedly that the GRB radiation is to be collimated, but the
collimation (2) concerns mainly only a small part of hard quanta.
The pairs production threshold for such quanta naturally and smoothly, 
according to the law $(1-cos\theta)^{1/2}$, rises 
with the decreasing of the angle between the direction at which the
photon is radiated from the surface of the compact object and
{\it a selected} direction (e.g. the magnetic field) on the surface. 
As a result, beside a soft component, 
the more and more hard part of the burst spectrum is passing through, 
and it is possible to suggest non-isotropic (axially symmetrical)
field of radiation around the source. 
Then, in particular, it is clear why 
XRF and XRR GRBs are uncollimated completely or rather almost isotropic 
(Lamb et al. 2003).

And what are typical sizes accepted in the standard model?
In the paper by Beloborodov (2004) namely the early stages of the GRB
explosion are considered. This is just the standard view:
GRB afterglow is explained as emission from
a decelerating blast wave. The deceleration begins at
$R_{dec} \sim 10^{15} - 10^{17}$\,cm.
It depends on the ambient density and initial Lorentz factor.
So, $R_{dec}$ is a fireball size before the deceleration begins.
And $R_{dec}$ does not exceed $10^{17}$\,cm.
In this theory this is actually the size of the region where the
GRB prompt emission (= \g-ray burst) with the observed spectrum
arises: $\sim 10^{15} - 10^{17}$\,cm. Further it is already a zone where the
{\it afterglow} arises --- hours and days after the prompt GRB.

But as was said above, allowing for the influence of the angle between
photon momenta
in the source on the threshold of $e^-e^+$ pair productions, 
it is possible to assume that the GRB radiation arises
in the region 10 orders smaller. But then
there must be quite different physical 
conditions providing the GRB source energy release.

\section{Possible mechanisms of GRB phenomenon in the compact model}

Certainly, after all, one should think about the main thing:
to assume and investigate a physical mechanism explaining
the GRB origin on the surface or close to 
an object of type of
neutron star (NS) or quark/strange star.
All possible versions of energy output onto the surface of a compact object or
of
explosions related somehow to this object should be considered,
see the reviews by Bisnovatyi-Kogan (2003, 2004).
Below we are examining in outline some of the compact mechanisms
of the GRB energy release.

A mechanism of the GRB origin in the vicinity of a collapsing object
based on neutrino-antineutrino annihilation was analyzed by
Berezinsky and Prilutsky (1987).
Earlier GRB production in a SN explosion under the action of neutrino
pulse was suggested by Bisnovatyi-Kogan et al. (1975).
The starquake, subsequent explosion and outburst from a non-equilibrium layer in
the neutron star crust, discovered by Bisnovatyi-Kogan
and Chechetkin (1974), is accompanied by gamma radiation due to fission of the
ejected super heavy nuclei. This scenario was suggested (Bisnovatyi-Kogan
et al. 1975) as an alternative model for GRBs of galactic origin,
but now it can be claimed again in the compact GRB model under consideration.
As was obtained by Brezinsky and Prilutsky (1987), the efficiency of
transformation of the neutrino flux energy $W_{\nu} \sim 6\cdot10^{53}$\,ergs
into X-ray and \g-ray burst is $\alpha \sim 6\cdot10^{-6}$, with the
energy output in GRB $W_{X, \gamma} \sim 3\cdot10^{48}$\,ergs.
This is in agreement
with the total energy of the GRB source of $10^{47} - 10^{49}$\,ergs
in the compact GRB model,
with accounts for strong beaming (2).

One of quite well developed mechanisms of compact energy output
in the vicinity of the collapsing object
(magnetorotational explosion) was suggested for the SN explosion
(Bisnovatyi-Kogan 1971).
Numerical calculations of the explosion in magnetized rotating gas
could give an efficiency of transformation of the rotational energy
into kinetic  at a level of 10 percent
(Ardeljan et al. 2000).
Some results of 2-D calculations of the magnetorotational explosion
produced by rapidly and differentially rotating, strongly magnetized
new-born NS are given by Moiseenko et al. (2003).
The energy output is sufficient for the SN explosion,
but seems to be low for the total ``standard" X-ray and \g-ray energy release of
$\sim 10^{51} - 10^{53}$\,ergs in the standard fireball GRB model.

Thus, the very idea  of existence of strong global magnetic fields in the
region of cosmic GRBs generation has already been given many times
(see also papers by Usov 1994; Thompson 1994; Meszaros and Rees 1997;
Blandford 2002).
In this connection another mechanism of  compact energy output or of the origin
of cosmic GRBs resulting from a decay of magnetized
vacuum around NS with such a field can be suggested.
Here the energy of order $10^{47} - 10^{49}$\,ergs
for the source of GRB in the compact GRB model
corresponds well to the value of vacuum energy near NS with
a super strong field
$B \sim 10^{15}\div 10^{16}$\,G on condition that the star
surface undergoes oscillations (Gnedin 2004).
One can suppose that such oscillations (or starquakes) occur
at the moment when the NS (or a new-born NS) is being formed as a result
of the massive core collapse of a star progenitor.
The value of the vacuum energy released in this case depends on both amplitude
and frequency of such oscillations.
A possibility of decay of the vacuum in a
super-strong magnetic field is now discussed rather actively
 (Calucci 1999; Xue 2003; Rojas and Querts 2004;
 Metalidis and Bruno 2003).
The idea of the decay of the strongly magnetized vacuum in the vicinity of NS
with a super-strong magnetic field for explanation
of the phenomenon of cosmic GRB was first stated
by Gnedin and Kiikov (2001), and the first
energy estimates were also obtained there.
The energy accumulated in strongly magnetized vacuum is quite sufficient
to provide the GRBs energy output in the process of decay of such vacuum.
Here the proper oscillations of the NS surface can be a triggering mechanism
of such a decay.
The result can  be a perfect realization of the following chain:
the collapse of a massive star core --- oscillations of the surface
of a new-born NS --- the collapse of magnetized vacuum with the energy output.
But the main question remains: how the energy of magnetized vacuum is
transformed into radiation? One of the possible solutions of this problem
based on analogy with the phenomenon of sonoluminescence is suggested
in the paper by Gnedin and Kiikov (2001). But the probability of such a process
demands a  separate special consideration (Gnedin 2004).

Two giant flares were observed on  March 5, 1979 and  August 27, 1998 from
(so-called) soft \g-ray repeaters SGR 0526-66 and SGR 1900+14, respectively.
The peak luminosity of these flares was as high as
$\sim 10^{45}$\,erg\,s$^{-1}$
(Mazets \& Golenetskii 1987; Hurley et al. 1999).
It will be recalled that the problem of GRB source compactness arose
just in the explanation of
the 1979 March 5 event (Carrigan \& Katz 1992)
with the huge luminosity
and with the very short observed rise time (to $10^{-3}$\,s).
Such a bursting activity of the SGRs can be explained by the fast heating
of the bare surface of a strange star and its subsequent
{\it thermal emission} (Usov 2001).
The heating mechanism may be, for example, the fast decay of super-strong
($\sim 10^{15} - 10^{16}$\,G) magnetic fields.
The energy output mechanism in this Usov's model
can be advantageously used to explain the long-duration (cosmological) GRBs
 in the compact model with the small collimation
of the prompt GRB radiation with strong beaming (2) for \g-rays.
In this case, GRBs should be considered as a set of short bursts
(like the giant flares of the 1979 March 5 and 1998 August 27 events)
with a total GRB duration of $\sim 10^2 - 10^3$\,s.

From the said above in this section it is seen that the attempts
to explain GRBs by physics related with a massive compact object
have  already a rather long history. This experience can be used for
detailed development of the compact model of GRB source
or GRB scenario with the compact energy output allowing for what
was said in the previous sections of our paper about observational
and theoretical arguments in favor of such an attempt to solve
the problem of the GRB source compactness.

\section{Concluding remarks}

So, XRFs can be not collimated at all or slightly collimated (XRR GRB),
but with the low total bolometric energy of $\sim 10^{47}$ ergs.
Since  most probably these are actually the explosions of massive SNe
at distances of 100 Mpc (Norris 2003; Norris \& Bonnel 2003),
they can be observed much more frequently than it is predicted by the standard
fireball GRB model.
One should try to find early spectral and photometrical SN features.
Then, in general,
the observational problem of XRF/XRR/GRB identification becomes a
special section in the study of cosmological SNe.
(It will be recalled that the GRB~030329/SN 2003dh was a XRR GRB but not
a classical GRB.)

As to normal/classical GRBs and especially those ones with many heavy
quanta in spectra, it is possible to obtain directly from formula (1) a
(kinematical) estimate of the limit collimation of this gamma
radiation, which, in turn, independently agrees with the observational ratio
(2) of the yearly rates
$N_{GRB}/N_{SN} \sim 10^{-5} - 10^{-6}$.
If the matter concerns quanta with $E\sim 100MeV$ of distant and the most
distant GRBs, then from
$1 - cos\theta_{12}\approx 0.5MeV^2/(100MeV\cdot100MeV) = 0.5\cdot 10^{-4}$
it follows that the radiation of such GRBs turns out to be the most
collimated.
Such photons must be radiated in the cone of an opening of $\approx 0.5^o$
and be detected in the spectra of the rather distant GRBs with $z\sim 1$ and
 farther because of geometrical factor only.

Thus, a natural consequence of our compact model of the GRB source 
is the fact
that distant bursts ($z\gtrsim 1$)
turn out to be harder ones, while close ``GRBs" ($z \sim 0.1$)
look like XRF and XRR GRBs with predominance of soft X-ray quanta 
in their spectra (though the factor $1+z$ also works).
Naturally, the effects of observational selection due to finite sensitivity
of GRB detectors should be also taken into account.
For example, the soft spectral component of the distant (classical) GRBs is
``cut" out by the detector sensitivity threshold.
And the isotropic X-ray burst, simultaneous with the GRB, can be simply not
seen in distant (classical) GRBs because of 
the  low total/bolometric luminosity of the source
in the compact GRB model ($< 10^{49}$ ergs). 
Actually, XRF and XRR GRBs have lower values of
$E_{iso}$ (so called isotropic equivalent),
than GRBs (Amati et al. 2002; Lamb et al. 2003a, 2003b, 2003c).

As a result, only radiation within a narrow solid angle near a selected
direction (in the GRB source) 
is observable for GRB detectors:
the soft spectral range remainder 
for the prompt XRR GRBs with $z\lesssim 1$
(having climbed over the equipment threshold)
and hard (and even heavy) quanta below the threshold 
of the pair production (1) for classical GRBs with $z\gtrsim 1$.
(Close XRFs with $z\lesssim 0.1$ can not have the $\gamma$-ray 
quanta in their spectra at all for the definit equipment threshold).
Though from the review by Postnov (1999) it follows that ``typical" GRBs
are seen in the range of 30 keV --- 100 MeV,
but it turns out (and it was known before) that  most GRBs are much
softer (see Sect. 2). 
Not without reason Lamb et al. (2003)
were just amazed by this important {\it observational}
result of BeppoSAX and HETE-2 missions.
We mean the detection of obvious XRFs and XRR GRBs first
by BeppoSAX (Amati et al. 2002) and then by HETE-2.
In our compact model of GRB source it (the Amati law)
can be a ``simple" consequence of formula (1) + collimation (most
probably) by magnetic field on the surface of the compact object.

In the scenario of jet formation, which was discussed in this paper and which
was also used to interpret the GRB 970508 optical transient (OT) light
curves (Sokolov 2001b) an isotropic X-ray, optical and radio emission of the
{\it afterglow} of the GRB OT is possible.
In X-ray lines
 (Piro et al. 1999; Yoshida et al. 2001; Piro et al. 2000;
Antonelli et al. 2000)
it was so for sure.
At that an initial assumption was just a possibility of the small GRB
collimation (2), which follows from the comparison of the rates of
GRBs and SN explosions in distant galaxies.
It means that the close relation between GRBs and SNe
was taken as a basic assumption.
{\it All} long GRBs are always accompanied by SN explosions, which are
sometimes observed, and sometimes not 
(Sokolov 2001a, 2002).
In other words, the long GRB is the beginning of a massive
star collapse  
or the beginning of SN explosion, and GRBs must always be accompanied
by SN explosions (of Ib/c type or of {\it other} types of massive SNe).
Then in any case the total energy release at the burst in \g-rays can be
{\it not more} than the total energy released by any SN ($ <$ or $\sim
10^{49}$\,ergs) in all {\it electromagnetic} waves.
(It is interesting that the total energy release in X-ray emission
lines observed
with BeppoSAX, ASCA, Chandra for GRB 970508, GRB 970828,
  GRB 991216, GRB 000214 is of the same order --- see the collected data in
the paper by Ghisellini et al. (2002)
``Emission lines in GRBs constrain the total energy reservoir.")

But with so ``low" total energy of the GRB explosion
($\lesssim 10^{49}$ ergs) the only possibility
to see GRB at cosmological distances ($z\gtrsim 1$) 
is the detection of at least the most collimated part
of this energy ($1-10$\%) leaving the source within the solid angle of
$\Omega_{beam} \sim (10^{-5} - 10^{-6})\cdot 4\pi$.
The rest can be inaccessible for GRB {\it detectors} with
a limit sensitivity of $\sim 10^{-7} erg \cdot s^{-1}\cdot cm^{-2}$.
Certainly, it does not concern the 10~000 times more sensitive X-ray 
{\it telescopes} which were used to make sky surveys with the  
Ariel V, HEAO-1, Einstein satellites (Heise et al. 2001).
For limit sensitivity of $\sim 10^{-11} erg\cdot s^{-1}\cdot cm^{-2}$ in
the band of $0.2-3.5$ keV the X-ray observatory 
(Einstein) recorded Fast X-ray Trasients
(unidentified with anything) at a rate of $\sim 10^6 yr^{-1}$ 
all over the sky.
It agrees well with an average rate of the massive SNe explosions 
in distant galaxies, 
but for the present, GRB-detectors see only $\sim 10^{-4}$
part of this huge
number of the distant SN explosions as GRBs.

It is natural that at the total/bolometric energy of ``GRB"
$\sim 10^{47} - 10^{49}$\,ergs and at the GRB energy (3) released in 
the narrow cone (2), ``the fireball" also looks in quite a different way.
As to the compactness problem solved by the fireball model
for GRB energies of $10^{52} - 10^{53}$\,ergs, there is no such a
problem for ``\g-burst" energies $\sim 10^{47} - 10^{49}$\,ergs.
In any case, allowing for the low \g-ray collimation from the surface
of the compact object -- 
GRB/XRR/XRF source, which is necessary for the
angular dependence of $e^-e^+$ pair production (1), this problem is
solved under quite different physical conditions 
in the GRB-source than that supposed by Piran (1999).
In the scenario:
{\it massive star} ---$>$ {\it WR} ---$>$ {\it pre-SN $=$ pre-GRB},
in which only a small part of the most collimated radiation with the
collimation (2) goes to infinity and, correspondingly, with the
total energy of $10^4 - 10^6$ times less than in the standard theory, the
source can actually be of a size $\lesssim 10^8$\,cm.
It means that at the energies of up to $\sim 10^{49}$\,ergs the old
(``naive") estimate of the source size resulting directly from the time
variability of GRB can be quite true.
Thus, the point can be that the burst energy is much less than in the
standard fireball model.

To the above-said we add that the suggested compact GRB scenario allows also
predicting the behavior of superluminal radio components (which, e.g., have
been observed recently for GRB 030329 (Taylor et al. 2004)). As we
discussed in Sect. 5, most likely there is 
no considerable deceleration of the jet/bullet
(with the Lorentz factor of order 10).
Hence we expect that the
superluminal radio components related to the jet have the following
properties: \\
1) the radio component will move with the constant observed superluminal
velocity; \\
2) the characteristic
observed velocity of the superluminal component is of the order of the Lorentz
factor, i.e. of order $10 c$.

{\bf Acknowledgements.} 
Authors are sincerely grateful to Yu. A. Shibanov for fruitful
discussions, T. N. Sokolova for the manuscript editing.
The work was supported by
RFBR grants: 04-02-16300, 01-02-17106, 03-02-17223, Program of
the Presidium of RAN ``Non-stationary Phenomena in Astronomy",
Program of Russian Science and Education Ministry.

\end{document}